\documentclass[aps,groupaddress,superscriptaddress,onecolumn,floatfix,amsmath,amssymb,graphicx]{revtex4-2}

\usepackage{graphicx}
\usepackage{comment}
\usepackage{float}
\usepackage{soul}
\usepackage{color}
\usepackage[T1]{fontenc}
\usepackage{bm}
\newcommand{\de}[1]{\text{d}#1 \;}
\newcommand{\mean}[1]{\langle #1 \rangle}
\usepackage{ulem}
\usepackage{lineno}

\begin{document}

\preprint{APS/123-QED}

\title{Large system population dynamics with non-Gaussian interactions}

\author{Sandro Azaele}
\affiliation{Laboratory of Interdisciplinary Physics, Department of Physics and Astronomy ``G. Galilei'', University of Padova, Padova, Italy}
\affiliation{INFN, Sezione di Padova, via Marzolo 8, Padova, Italy - 35131}
\affiliation{National Biodiversity Future Center, Piazza Marina 61, 90133 Palermo, Italy}

\author{Amos Maritan}
\affiliation{Laboratory of Interdisciplinary Physics, Department of Physics and Astronomy ``G. Galilei'', University of Padova, Padova, Italy}
\affiliation{INFN, Sezione di Padova, via Marzolo 8, Padova, Italy - 35131}
\affiliation{National Biodiversity Future Center, Piazza Marina 61, 90133 Palermo, Italy}

\begin{abstract}
\textbf{Abstract:} 
We investigate the Generalized Lotka-Volterra (GLV) equations, a central model in theoretical ecology, where species interactions are assumed to be fixed over time and heterogeneous (quenched noise). Recent studies have suggested that the stability properties and abundance distributions of large disordered GLV systems depend, in the simplest scenario, solely on the mean and variance of the distribution of species interactions. However, empirical communities deviate from this level of universality.

In this article, we present a generalized version of the dynamical mean field theory for non-Gaussian interactions that can be applied to various models, including the GLV equations. Our results show that the generalized mean field equations have solutions which depend on all cumulants of the distribution of species interactions, leading to a breakdown of universality. We leverage on this informative breakdown to extract microscopic interaction details from the macroscopic distribution of densities which are in agreement with empirical data. Specifically, in the case of sparse interactions, which we analytically investigate, we establish a simple relationship between the distribution of interactions and the distribution of species population densities.
\end{abstract}

\maketitle

\section{Introduction}\label{sec1}

The exploration of the links between the microscopic characteristics of physical systems and their (macroscopic) collective behaviour dates back to the dawn of statistical mechanics. In his doctoral thesis the Dutch physicist J.D. van der Waals showed how to understand the macroscopic phases of liquids and gases in a unified way \cite{van1873over}. By making certain assumptions about the finite size and the mutual attractive forces of particles, he explained that a fluid can exist in either of two different macroscopic states, which have different densities. 
He also showed that it is possible to switch from one to another through what we now dub a phase transition.

Since its inception, this approach has hugely increased its power and found applications in fields that are far from physics \cite{may1972will,diederich1989replicators,fisher2014transition}. Today we have tools to understand systems where individual entities interact and produce large scale patterns. Ecology makes no exception and offers further challenges which are not present in the physics context \cite{Kerner1957,Goel1971,azaele2016statistical,Peruzzo2020,pearce2020stabilization,Roy2020}. Also, the rise of next-generation sequencing techniques has resulted in a growing wealth of ecologically significant data. These data are instrumental in characterizing microbial communities across various environments and involving a vast number of species \cite{integrative2019integrative}. However, despite such a deluge of empirical data, our understanding of the underlying mechanisms and functioning of these biological systems remains far from complete. Indeed, we have not understood yet the fundamental equations that describe the evolution of an ecosystem under generic conditions, and usually the existence of a Hamiltonian or constants of motion is a sign that a model may introduce some unrealistic biological features \cite{Kerner1957,Goel1971}. Thus, although ecological systems cannot be thought of as physical systems in thermal equilibrium, they anyway possess general laws \cite{Lawton1999}.

One of the emergent patterns is the distribution of population densities across species (SAD or RSA) \cite{azaele2016statistical,Bowler2012}. This informs us about how many species are rare with only a few individuals and how many are widespread and common. The SAD of highly diverse communities is quite uneven, but the most remarkable feature is its apparent universality \cite{Volkov2007,Chisholm2010}: despite major ecological differences between ecosystems, the shape of the distribution is remarkably similar. The SAD of coral reefs, tropical forests, breeding birds or even bacterial communities can often be described by curves that are well approximated by gamma-like distributions \cite{Volkov2007,grilli2020macroecological}.

What is the nature of this universality \cite{colyvandawn}? Does it reflect fundamental laws of community assembly or is that simply an inescapable statistical pattern which does not really help understanding how ecosystems work \cite{McGill2007}? Archetypal structures of interactions, including competition, mutualism or parasitism, may guide the search for the underpinnings of coarse-grained properties. Emergent patterns may also be the result of a bottom-up control of distribution of resources among individuals. However, one may conjecture simpler explanations: species have different characteristics and have evolved a variety of traits, but none of them play a central role in the ecosystem assembly and individuals interact with no specific structure. This ‘heterogeneity without order’ hypothesis could underpin generic features of ecosystems, including the variation of species’ abundances \cite{barbier2018generic}.

Whilst the shortcomings of this parsimonious assumption are well-recognized, it is also clear its simplicity: ecosystem assemblages are not primarily shaped by regular motifs in the interactions of individuals or by species’ traits which evolution has honed for coexistence. More simply, ecosystems are wrought by a myriad of tiny biological differences, which cease to matter as the number of interacting species becomes very large. 

The violation of this paradigm is informative as well. For instance, deviations from idealized disordered ecosystems are instrumental to gaining deeper insights into how species specifically interact, thus unveiling essential features of real ecosystems. This has the potential to untie the links between coarse-grained patterns, such as the SAD, and fine scale individual interactions. But how realistic are the universal patterns ensued from interactions that are dominated by randomness?

Disordered materials have been the focus of statistical physics for a long time \cite{mezard1987spin}. Despite  disordered systems have originated in Physics \cite{sompolinsky1981dynamic,opper1992phase,angelini2017real}, its notion and tools have reached a wealth of different fields, including neural networks \cite{sompolinsky1988chaos,mazzucato2023baseline}, polymers \cite{Montanari04poly}, systemic risk in economy \cite{paga2015contagion} and information theory \cite{mezard2009information}. More specifically, the dynamical mean-field theory (DMFT) has proven a useful theoretical tool to investigate the effects of quenched disorder in dynamical systems where the number of interacting degrees of freedom is very large. More recently, DMFT has been applied to large ecosystems driven by generalized Lotka-Volterra equations, in which the interactions between species are commonly modeled by taking random couplings \cite{bunin2017ecological,tikhonov2017collective,biroli2018marginally,galla2018dynamically,roy2020complex,altieri2021properties}. The approach has mainly focused on the stability-complexity properties of the ensuing disordered communities, along with the general features of their phase diagrams. So far, however, it has been difficult to provide convincing predictions of realistic community patterns. Therefore, although the DMFT represents the natural generalization of the conceptual framework pioneered by May in his seminal 1972 paper \cite{may1972will}, current applications to community ecology present some challenging aspects.

In this article we depart from the interests in complexity-stability issues of recent literature. Instead, we harness the power of disordered systems techniques to dig deep into the relations between microscopic interactions and emergent community patterns in random ecosystems. This will require a genuinely new version of the DMFT which will give access to completely new properties of the framework. The outcome will be the prediction of patterns that are closer to those observed in empirical ecosystems, with a broad insight into how species interact.
 
\section{Results}\label{sec2}

\subsection{General Setting}\label{subsec1}
We consider an ecosystem consisting of $N$ species, whose density of individuals at time $t$ is $x_i(t)$ for any $i=1,\ldots,N$. These variables are continuous, positive and evolve according to the generalized Lotka-Volterra (GLV) equations

\begin{equation}
    \dot{x}_i(t)=x_i(t)\left(r_i+\sum_{j=1}^{N}\alpha_{ij}x_j(t)\right)\quad,
    \label{eq:glv}
\end{equation}
where $r_i$ are the intrinsic growth rates of species. In the following we will set $r_i=1$ and $\alpha_{ii}=-1$ for the sake of argument, but the applicability of our method does not depend on this assumption. In this model species are driven by pairwise interactions indicated by the coefficients $\alpha_{ij}$, which specify the per capita effect between different species' population densities (for $i\neq j$) and in general are not symmetric. With our notation, a negative coefficient $\alpha_{ij}$ indicates that the species $j$ is deleterious to the species $i$ because of competition or predation.

In the standard setting of dynamical mean field theory (DMFT) applied to ecosystems the coefficients $\alpha_{ij}$ are identical and independent random variables which are drawn from a Gaussian distribution defined by its mean and covariance matrix. Actually, the exact form of the distribution of the coefficients is not important as long as its mean and variance exist, both decrease with $N^{-1}$ and all higher cumulants decay faster than $N^{-1}$ in the limit of large $N$. This specific scaling ensures that the `thermodynamic' limit may produce systems with a large number of coexisting species. 

This would be a strong result if the patterns predicted by the DMFT were in agreement with the real ones. However, the SAD predicted by the standard DMFT is usually a truncated Gaussian-like distribution -- at least in the theoretically controlled regimes of the theory --, and therefore rarely in agreement with the empirical data. In addition to this, the theory teaches us the following important lesson. In its simplest form the model defined by eq.(\ref{eq:glv}) is characterised by two parameters only (the mean and variance of interaction strengths between species) in the large $N$ limit. As a consequence, the SAD of disordered ecosystems becomes a universal pattern which does not depend on the details of the specific species' interactions. This means that the SAD bears little resemblance of particular structures in the interactions as it is essentially shaped by their mean and variance. Therefore, we cannot trace them back from the SAD, simply because there is not enough information for that in the SAD.

These considerations call for a generalization of the framework which is able to keep the advantages, whilst reaching a more realistic description of ecosystems. Instead of considering only the first two cumulants as alluded to above, we introduce a distribution of the off-diagonal coefficients $\alpha_{ij}$, $P_N(\alpha)$, whose characteristic function scales with $N$ as

\begin{equation}
\lim_{N\to +\infty}N \ln \left( \int \text{d}\alpha\; P_N(\alpha)e^{-i\alpha z }\right)= F(z) \quad, 
    \label{eq:scaling inter}
\end{equation}
where $F(z)$ is a complex function which is analytic at least at $z=0$, and $F(0)=0$ to ensure the normalization of $P_N(\alpha)$ for any $N$. Notice that (when $F(z)$ is not a polynomial) eq.(\ref{eq:scaling inter}) implies that \textit{all} cumulants of $P_N(\alpha)$ are of order $N^{-1}$ in the limit of large $N$, because the characteristic function of a probability density function is the exponential of its cumulant-generating function (see the Supplementary Information (SI); also, if we want that $\exp(F(z)/N)$ is the generating function of a PDF, $F(z)$ is either at most a polynomial of degree two or an infinite power series -- see Marcinkiewicz's theorem in \cite{lukacs1960characteristic}). When $F(z)=-i \mu z - \sigma^2 z^2/2$ we recover the assumptions of the standard DMFT, for which $\lim_N N \mean{\alpha_{ij}}=\mu$ and $\lim_N N (\mean{\alpha_{ij}^2}-\mean{\alpha_{ij}}^2)=\sigma^2$. 

The condition specified in Eq.(\ref{eq:scaling inter}) encapsulates more information of the microscopic distribution of species' interactions with respect to the standard approach with Gaussian couplings, and provides the theory with more flexibility when the ecosystem becomes very biodiverse. This is central to create a link between the distribution of the coefficients $\alpha_{ij}$ and the macroscopic pattern of the SAD. 

Let us consider a paradigmatic example. For the sake of simplicity, we draw the coefficients $\alpha_{ij}$ of eq.(\ref{eq:glv}) from the gamma-like distribution

\begin{equation}
P_N(\alpha)= \frac{\lvert \alpha-\mu/N \rvert ^{-1+\Bar{\delta}(N)}}{2 \Bar{\beta}(N)^{\Bar{\delta}(N)}\Gamma(\Bar{\delta}(N))}e^{-\lvert \alpha-\mu/N \rvert /\Bar{\beta}(N)}\quad,
    \label{distr:gamma}
\end{equation}
where $\mean{\alpha_{ij}}=\mu/N$, $\Gamma(x)$ is the gamma function and $\Bar{\delta}(N)$ and $\Bar{\beta}(N)$ are two positive functions of $N$ which we are going to specify according to an appropriate large $N$ limit. If we choose $\Bar{\delta}(N)=\delta$ and $\Bar{\beta}(N)=\beta/\sqrt{N}$ we obtain $F(z)=-i\mu z-\beta^2\delta(\delta+1)z^2/2$, thus recovering the standard results of the DMFT for $\sigma^2=\beta^2\delta(\delta+1)$. However, for the case $\Bar{\delta}(N)=\delta/N$ and $\Bar{\beta}(N)=\beta$ we obtain

\begin{equation}
F(z)=-i\mu z-\frac{\delta}{2}\log(1+\beta^2 z^2)\quad,
    \label{f:effedizeta}
\end{equation}
which is different from the standard case and introduces new features which we will investigate in the following sections.

\subsection{The mean-field equations}
We will investigate the properties of the random GLV introduced in Eq.(\ref{eq:glv}) using methods from statistical physics of disordered systems. The key theoretical device which allows to study the effects of quenched noise in the large $N$ limit is the dynamical generating functional associated to the system (\ref{eq:glv}). This is given by

\begin{equation}
Z[\vec{\psi}] = \int  D[\vec{x}, \vec{\tilde{x}}]\exp \left\{S[\vec{x}, \vec{\tilde{x}}] + i\sum_{j=1}^N \int dt \, x_j(t) \psi_j(t)\right\} \quad,
    \label{eq:gen func}
\end{equation}
where $\psi_j(t)$ represent external source fields that are used to calculate correlations, $D[\vec{x}, \vec{\tilde{x}}]$ indicates a functional integration over the solutions of Eq.(\ref{eq:glv}) and $S[\vec{x}, \vec{\tilde{x}}]$ is the Martin-Siggia-Rose-DeDominicis action \cite{martin1973statistical,janssen1976lagrangean,de1978field,chow2015path} of the system which is written down explicitly in the SI, along with other details of the calculations. This functional weighs all possible trajectories generated by the GLV dynamics with Fourier weights and can be used to calculate some key quantities which characterize the dynamics of the system. 

Finding the average over all possible realizations generated by the heterogeneity of the interactions, using the condition in Eq.(\ref{eq:scaling inter}) and finally taking the limit $N\to +\infty$, we obtain the following stochastic differential equation for the density of the population size, $x(t)$, of a representative species 

\begin{equation}
\dot{x}(t)=x(t)[1-x(t)+\eta(t)]\quad,
    \label{eq:langevin}
\end{equation}
where 
$\eta(t)$ is a stochastic process whose cumulants are defined in terms of the moments of the stochastic process $x(t)$, namely, for any $r=1,2,\ldots$

\begin{equation}
\mean{\prod_{k=1}^r\eta(t_k)}_C=i^r b_r r! \mean{\prod_{k=1}^r x(t_k)}\quad,
    \label{eq:cumulants time}
\end{equation}
where $b_r$ is the $r$-th coefficient of the power series $F(z)=\sum_{r>0}b_rz^r$. Here $\mean{\cdots}$ indicates the average with respect to the noise $\eta(t)$ and $\mean{\cdots}_C$ indicates a cumulant; thus, for $r=1,2$, $\mean{\eta(t)}=\mu \mean{x(t)}$ and $\mean{\eta(t_1)\eta(t_2)}-\mean{\eta(t_1)}\mean{\eta(t_2)}=\sigma^2 \mean{x(t_1)x(t_2)}$. We now describe the principal aspects of the process defined in Eqs.(\ref{eq:langevin}) and (\ref{eq:cumulants time}). More technical details can be found in the SI.

The conditions in Eq.(\ref{eq:cumulants time}), which constrain the cumulants of $\eta(t)$ to be proportional to the moments of $x(t)$, imply that Eq.(\ref{eq:langevin}) must be solved in a self-consistent way at any time $t>0$. Later on, we will discuss the main properties of the stationary distribution. The Langevin equation (\ref{eq:langevin}) was deduced by assuming that different coefficients $\alpha_{ij}$ are not correlated; this is crucial for obtaining an equation that is local in time, which leads to a Markovian dynamics. Interestingly, this equation has the same form as the one derived in the standard DMFT \cite{bunin2017ecological,galla2018dynamically}. However, the current version of the theory introduces a new crucial property of the noise: whilst in the original framework $\eta(t)$ was a coloured Gaussian noise with self-consistent temporal correlations, in this more general setting $\eta(t)$ is a coloured non-Gaussian noise, whose cumulants depend on all $n$-point temporal correlations in a self-consistent fashion.\\

\begin{table}[ht]
 \centering
\begin{tabular}{|p{2cm}|p{0.3cm}|p{0.3cm}|p{3cm}|p{2cm}|p{2cm}|}
 \hline
 \multicolumn{6}{|c|}{Distribution of interactions: $P_N(\alpha)=\dfrac{\lvert \alpha-\mu/N\rvert^{-1+\delta/N}\exp{\left(-\lvert\alpha-\mu/N\rvert/\beta\right)}}{2\beta^{\delta/N}\Gamma{\left(\delta/N\right)}}$} \\
 \hline
Parameters & $\mu$ & $\sigma$ & Order parameter & $\phi$ & $\phi\sigma^2$\\
 \hline
$\delta=1$, $\beta=1$& -3 & 1 & $6\cdot 10^{-3}$ ($2\cdot 10^{-3}$)& 0.89 (0.01) & 0.89 (0.01)\\
$\delta=\frac{1}{4}$, $\beta=2$& -3 & 1 & $3\cdot 10^{-3}$ ($2\cdot 10^{-3}$)& 0.94 (0.01) & 0.94 (0.01)\\
$\delta=1$, $\beta=2$& -3 & 2 & $5\cdot 10^{-2}$ ($8\cdot 10^{-2}$) &  0.74 (0.02)& 2.9 (0.1)\\
$\delta=4$, $\beta=1$& -3 & 2 & 0.1 (0.1) &  0.4 (0.1)& 1.5 (0.5)\\
$\delta=1$, $\beta=3$& -3 & 3 & 0.06 (0.07) &  0.6 (0.2)& 6 (2)\\
$\delta=9$, $\beta=1$& -3 & 3 & unbounded growth &  -- & --\\
$\delta=1$, $\beta=1$& -2 & 1 & $7\cdot 10^{-3}$ ($5\cdot 10^{-3}$) & 0.87 (0.02) & 0.87 (0.02)\\
$\delta=\frac{1}{4}$, $\beta=2$& -2 & 1 & $2\cdot 10^{-3}$ ($2\cdot 10^{-3}$) & 0.94 (0.01) & 0.94 (0.01)\\
$\delta=1$, $\beta=2$& -2 & 2 & $10^{-2}$ ($6\cdot10^{-3}$) &  0.77 (0.02)& 3.1 (0.1)\\
$\delta=4$, $\beta=1$& -2 & 2 & $0.1$ ($0.3$) &  0.3 (0.1)& 1.4 (0.4)\\
$\delta=1$, $\beta=3$& -2 & 3 & 0.1 (0.1)&  0.5 (0.3)& 4 (2)\\
$\delta=9$, $\beta=1$& -2 & 3 & unbounded growth &  -- & --\\
$\delta=1$, $\beta=1$& -1 & 1 &$4\cdot 10^{-3}$ ($3\cdot 10^{-3}$)& 0.88 (0.02) & 0.88 (0.02)\\
$\delta=\frac{1}{4}$, $\beta=2$& -1 & 1 &$3\cdot 10^{-3}$ ($3\cdot 10^{-3}$)& 0.94 (0.02) & 0.94 (0.02)\\
$\delta=1$, $\beta=2$& -1 & 2 &$10^{-2}$ ($6\cdot 10^{-3}$)& 0.7 (0.1) & 2.9 (0.6)\\
$\delta=4$, $\beta=1$& -1 & 2 & 0.1 (0.1) &  0.4 (0.1) & 1.7 (0.4)\\
$\delta=1$, $\beta=3$& -1 & 3 & 0.02 (0.02) &  0.5 (0.2) & 5 (2)\\
$\delta=9$, $\beta=1$& -1 & 3 & unbounded growth &  -- & --\\
$\delta=1$, $\beta=1$& 0 & 1 &$2\cdot 10^{-3}$ ($2\cdot 10^{-3}$)& 0.87 (0.02) & 0.87 (0.02)\\
$\delta=\frac{1}{4}$, $\beta=2$& 0 & 1 &$3\cdot 10^{-4}$ ($8\cdot 10^{-4}$)& 0.94 (0.02) & 0.94 (0.02)\\
$\delta=1$, $\beta=2$& 0 & 2 &$2\cdot 10^{-3}$ ($2\cdot 10^{-3}$)& 0.76 (0.03) & 3.1 (0.1)\\
$\delta=4$, $\beta=1$& 0 & 2 & unbounded growth &  -- & --\\
$\delta=1$, $\beta=3$& 0 & 3 &$3\cdot 10^{-3}$ ($7\cdot 10^{-3}$)& 0.67 (0.04) & 6.0 (0.4)\\
$\delta=9$, $\beta=1$& 0 & 3 & unbounded growth &  -- & --\\
 \hline
\end{tabular}

\caption{This table investigates numerically the phase diagram of the GLV ($N=200$) in Eq.(\ref{eq:glv}) when the interaction coefficients are drawn from the distribution $P_N(\alpha)$ defined on the top of the table. The limiting standard deviation is $\lim_N \sqrt{N}\bar{\sigma}(N)=\sigma=\beta \sqrt{\delta}$. The order parameter is the ratio of the standard deviation to the mean (Coefficient of Variation) of population densities at stationarity, averaged over $N$ species and across realizations. Relatively small values of the order parameter are compatible with systems with a unique equilibrium; `unbounded growth' indicates unstable numerical integrations which are compatible with unbounded growth of population densities. $\phi$ is the fraction of surviving species (population densities larger than $10^{-6}$). $\phi$ is unreliable when the order parameter is relatively large. Numbers within brackets indicate one standard deviation calculated over realizations.}
\label{table:1}
 \end{table}

\textbf{Stationary properties.} In this section we want to focus more closely on the relationships between the stationary properties of the processes $x(t)$ and $\eta(t)$. Henceforth, we will set $\mu=0$ for simplicity, but our results are not limited to this case. Assuming that $x(t)$ and $\eta(t)$ eventually converge to the respective fixed points, $x^*$ and $\eta^*$, independently of their initial conditions, the Langevin equation (\ref{eq:langevin}) admits two stationary states: $x^*=0$, which is the stable one if $1+\eta^*<0$ and 

\begin{equation}
x^*(\eta^*)= (1+\eta^*)\Theta (1+\eta^*)\quad,
    \label{eq:sol}
\end{equation}
when $1+\eta^*>0$, where $\Theta(x)$ is the Heaviside function, $\Theta(x)=1$ for $x>0$, and $\Theta(x)=0$ else. Therefore, after a transient time only a fraction of the initial species will survive. At stationarity, the self-consistent conditions in Eq.(\ref{eq:cumulants time}) read

\begin{equation}
\mean{(\eta^*)^r}_C=i^r b_r r! \mean{(x^*)^r}\quad,
    \label{eq:cumulants stat}
\end{equation}
and therefore $x^*$ is distributed like the process $1+\eta^*$ for $\eta^*>-1$, but in general they are not Gaussian-distributed, because of Eq.(\ref{eq:cumulants stat}). At this point we can connect the distribution of $x^*$, $Q_x(x)$, to that of $\eta^*$, $P_{\eta}(\eta)$, (we have removed the superscripts for semplicity of notation) by using a simple change of random variables:

\begin{equation}
Q_x(x)\equiv \mean{\delta(x-x(\eta))}_{\eta}=\delta(x)\int_{-\infty}^{-1}\de{\eta}P_{\eta}(\eta) + P_{\eta}(x-1)\Theta(x)\quad,
    \label{eq:x eta}
\end{equation}
where $x(\eta)$ is defined in Eq.(\ref{eq:sol}). The factor of the Dirac-delta quantifies the fraction of extinct species and in the following we will indicate it with $1-\phi$, being $\phi$ the fraction of surviving species. Notably, the distribution $Q_x(x)$ corresponds to the stationary SAD of the random GLV model in the large $N$ limit. This is the macroscopic pattern which can be observed in real ecosystems and is directly connected to $P_{\eta}(\eta)$, which encapsulates the information of the distribution of the interaction coefficients between pairs of species. The connection between $Q_x(x)$ and $P_{\eta}(\eta)$ expressed in Eq.(\ref{eq:x eta}) seems apparently useless, because the distribution of $\eta$ depends on that of $x$, according to Eq.(\ref{eq:cumulants stat}), which we actually wish to find. Instead, in the SI we show how to obtain a closed equation for $P_{\eta}(\eta)$ which depends only on $F(z)$. The equation is 

\begin{equation}
P_{\eta}(\eta) = \int_\mathbb{R} \frac{dz}{2 \pi} \exp\left\{i z \eta + \int_{-1} ^ \infty d \eta' P_\eta(\eta') F(z+ z \eta')\right\} \quad.
    \label{eq:distr eta}
\end{equation}
This is a non-linear, integral equation for the distribution $P_{\eta}(\eta)$ which allows us to link the microscopic interaction of species to the macroscopic pattern of the SAD.\\

\textbf{Properties of the noise.} The integral equation defined in Eq.(\ref{eq:distr eta}) is difficult to solve and we have not been able to find a general solution. Nevertheless, a few properties can be deduced. When $F(z)$ has the form of an infinite power series, $P_{\eta}(\eta)$ must decay faster than any power of $\eta$ (at least for large positive $\eta$), otherwise the integral in the exponent of Eq.(\ref{eq:distr eta}) would not exist. The moments of the distribution $P_{\eta}(\eta)$ are therefore finite. Also, when $F(z)$ is a second degree polynomial, Eq.(\ref{eq:distr eta}) can be solved analytically and the solution is a Gaussian distribution, whose conditions for the mean and variance are equivalent to the ones found in \cite{bunin2017ecological,galla2018dynamically}. However, a Gaussian \textit{ansatz} does not work for different forms of $F(z)$, which shows that a solution cannot be found by simply invoking the central limit theorem. Thus, Eq.(\ref{eq:distr eta}) is a genuine generalization of the standard DMFT when a solution exists. 

\begin{figure}[h!]
    \centering
\includegraphics[width=10cm]{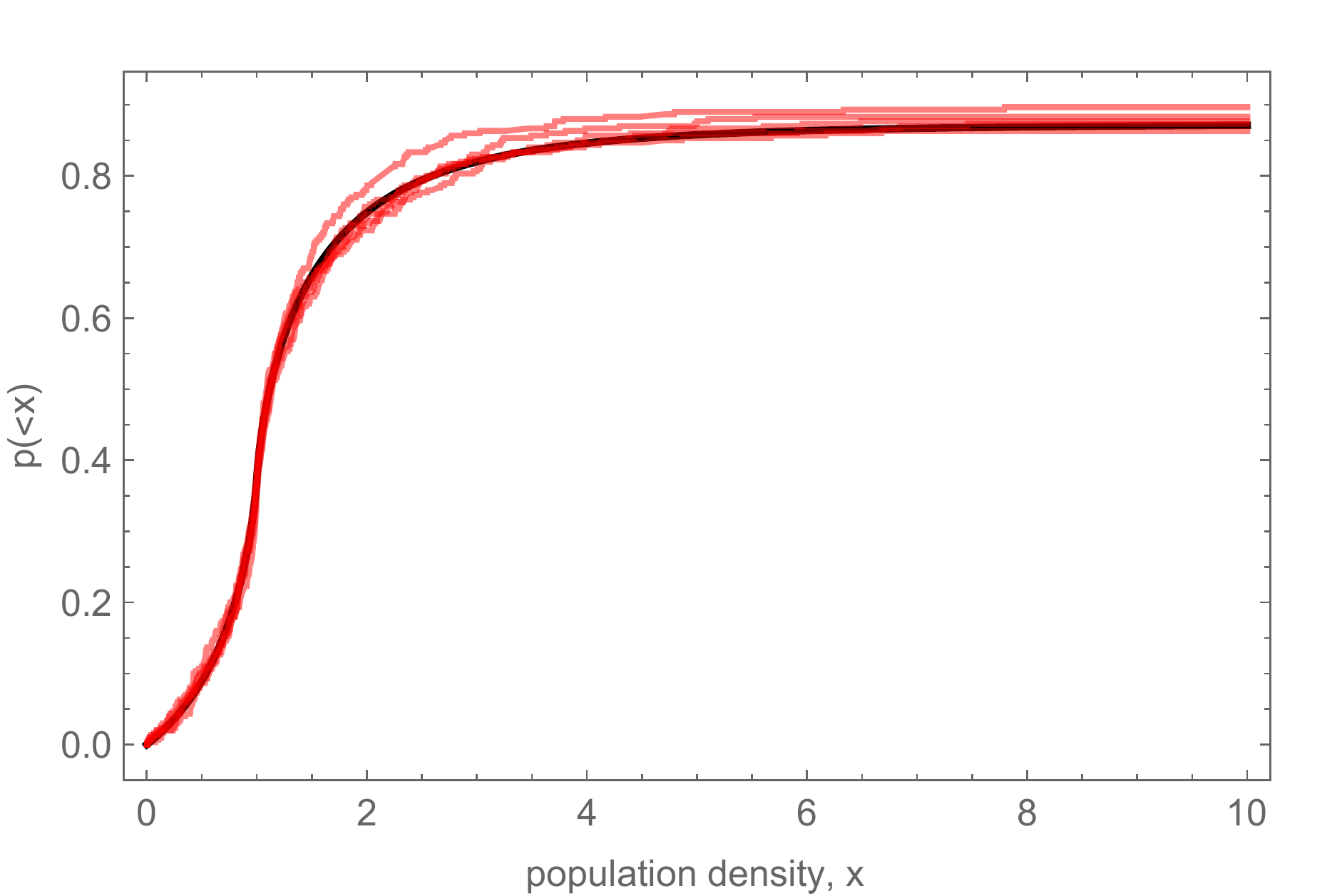}
    \caption{The black solid curve represents the fraction of species with a population density smaller than $x$ (this is the cumulative distribution of the Relative Species Abundance). It was obtained from the numerical solution of Eq.(\ref{eq:distr eta}) combined with Eq.(\ref{eq:x eta}). The jittery red lines come from a few realizations of the GLV in Eq.(\ref{eq:glv}) with $N=300$ species whose interaction coefficients $\alpha_{ij}$ are drawn from the distribution defined in Eq.(\ref{distr:gamma}), where $\lim_N N\bar{\delta}(N) =\delta= 1$; $\bar{\beta}(N) = \beta=1$; $\lim_N \sqrt{N}\bar{\sigma}(N)=\beta\sqrt{\delta}=1$ and $\mu=0$. The GLV system was run until stationarity, starting from uniform initial conditions $U[10, 20]$ and diagonal elements $\alpha_{ii}=-1$. 
    The plateau reached by the curves for relatively large population densities gives the fraction of surviving species ($\phi=0.87$). 
    }
    \label{fig:cum-rsa}
\end{figure}

We studied the case in which the interaction coefficients $\alpha_{ij}$ were drawn from the distribution defined in Eq.(\ref{distr:gamma}), which corresponds to solving Eq.(\ref{eq:distr eta}) with the function $F(z)$ given in Eq.(\ref{f:effedizeta}). The solution was found numerically by using an iteration scheme. After a few iterations, an initial exponential distribution converged to a final distribution, which was robust against different initial conditions. We have then calculated the ensuing cumulative distribution of the relative species abundance (RSA), which we report in Fig.(\ref{fig:cum-rsa}).

\begin{figure}[h!]
    \centering
\includegraphics[width=10cm]{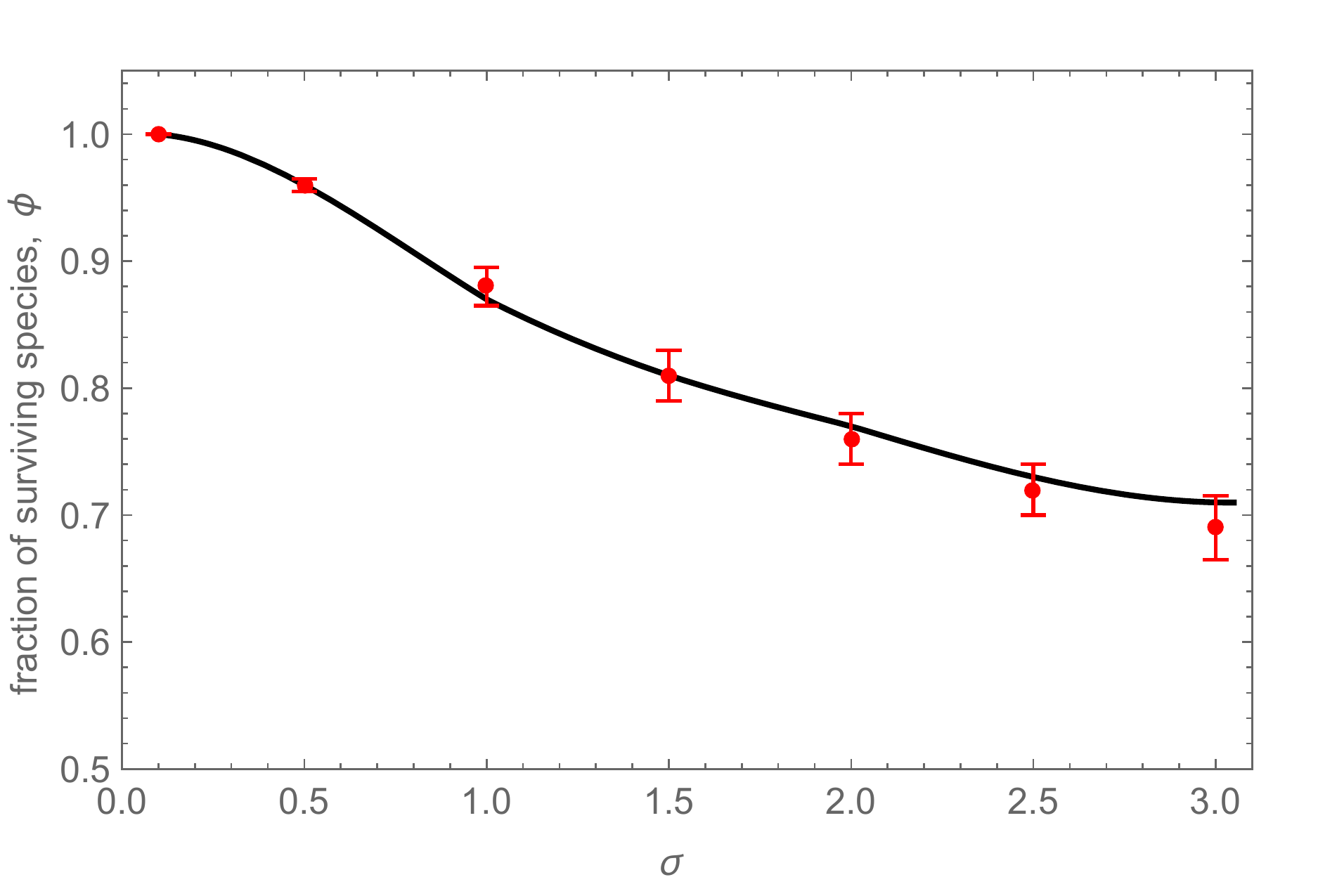}
    \caption{Fraction of surviving species, $\phi$ (see Eq.(\ref{eq:x eta})), as a function of the scaled standard deviation (i.e., $\lim_N \sqrt{N}\bar{\sigma}(N)=\sigma$). Red dots indicate averages from simulations of the GLV in Eq.(\ref{eq:glv}) with $N=100$, which were run until stationarity and whose interaction coefficients $\alpha_{ij}$ were drawn from the distribution defined in Eq.(\ref{distr:gamma}) with $\mu=0$.
    The average was zero and different scaled standard deviations were obtained 
    by varying $\beta$ only. Surviving species were those with population densities larger than $10^{-6}$. Error bars indicate one standard deviation calculated from 50 realizations of each system. The black solid line represents 
    the prediction obtained from the numerical integration of eq.(\ref{eq:distr eta}).}
    \label{fig:surv_species}
\end{figure}

The numerical integration of Eq.(\ref{eq:distr eta}) predicted a fraction of surviving species which was in very good agreement with the one obtained from the simulations of the GLV in Eq.(\ref{eq:glv}) (see Fig.(\ref{fig:surv_species})). Interaction coefficients were distributed according to Eq.(\ref{distr:gamma}), where we varied the parameter $\bar{\beta}(N)=\beta$ to obtain different values of the scaled standard deviation (i.e., $\lim_N \sqrt{N}\bar{\sigma}(N)$, where $\bar{\sigma}(N)$ is the standard deviation of the distribution defined in Eq.(\ref{distr:gamma})). The RSA has a general exponential decay for large population densities which in general is different from the Gaussian decay found in the standard DMFT. 

\begin{figure}
    \centering
\includegraphics[width=10cm]{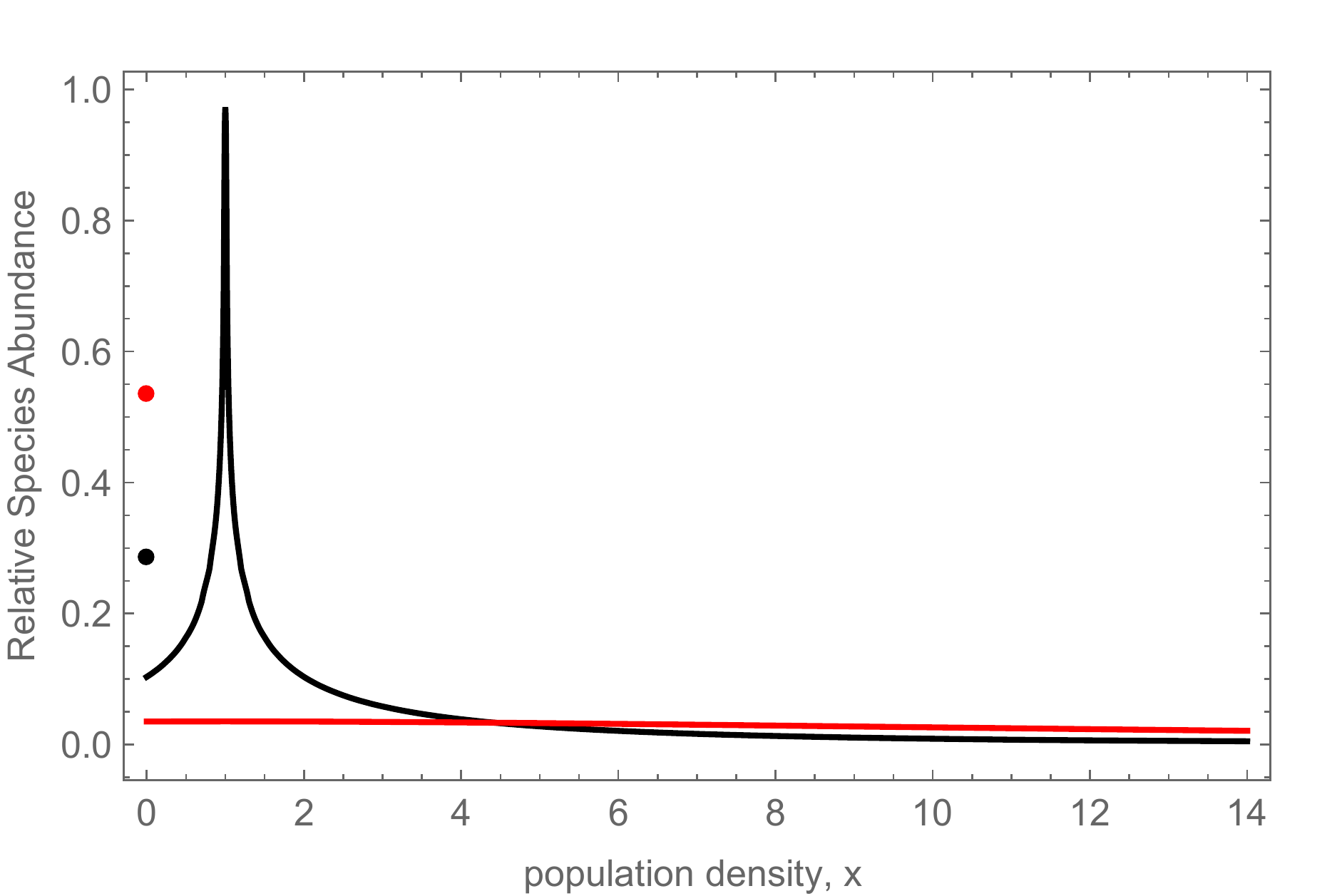}
    \caption{The Relative Species Abundance, i.e. the fraction of species with population density $x$, as predicted by the numerical integration of Eq.(\ref{eq:distr eta}) for the case $F(z)=-(\delta/2)\ln(\beta^2 z^2+1)$, which corresponds to a symmetric gamma-distribution of species' interactions with zero average. The parameters are $\lim_N N\bar{\delta}(N) =\delta$, $\beta$ does not scale with $N$ and $\lim_N \sqrt{N}\bar{\sigma}(N) = \sigma=\beta\sqrt{\delta}=3$. The black solid line was obtained for $\delta=1$, $\beta=3$, whereas the red solid line for $\delta=9$, $\beta=1$, leading to the same $\sigma$ ($\mu=0$) of the distribution of interactions. Black and red dots indicate the fraction of extinct species in the corresponding cases. The red curve indicates a non-normalizable solution and corresponds to an unbounded growth for the GLV system as reported in Table 1. It also shows that the generalized DMFT suggests a phase diagram which differs from the one of the standard DMFT.  
    }
    \label{fig:rsa_div}
\end{figure}

Finally, we have numerically verified that the phase diagram is different from the one we would obtain in the standard DMFT scenario. The most striking difference is the following: when the function $F(z)$ is a generic power series, the scaled mean and variance of the limiting distribution are not sufficient to define the regions of one equilibrium, multiple attractors or unbounded growth of the system. They depend on higher cumulants as well. This is in agreement with the predictions of Eq.(\ref{eq:distr eta}) as shown in Fig.(\ref{fig:rsa_div}). We have selected two sets of parameters which lead to a limiting distribution with the same mean and variance: the numerical integration of Eq.(\ref{eq:distr eta}) converged to a solution (black solid curve) only in one case, while in the other (red solid curve) the numerical solution was not normalizable. The dependence on higher cumulants of the phase diagram of Eq.(\ref{eq:glv}) is also confirmed by the simulations (see Table \ref{table:1}, in which we have also used different values of $\mu$).

\subsection{The case with sparse interactions}

Species that coexist in real world ecosystems rarely interact with all other species that are present in the community, unless this latter is very small. The matrix of the interaction coefficients $\alpha_{ij}$ that we have considered in the previous sections did not account for this possibility. This further level of realism leads to the introduction of another element of randomness, because one has to specify the number of species with which a given one interacts, and this is usually a random variable. Sparsity has proven challenging in the standard DMFT, but our framework allows analytic insights on the problem. 

We will examine two dilution regimes: one where a species interacts with only a few others, regardless of the total number of species in the system, and another regime where every species is almost connected to all the remaining ones, resulting in a system that is close to being fully connected. In this latter scenario, careful scaling of the interaction coefficients is necessary to restore the mean-field solution, which describes the non-diluted case mentioned earlier.

For describing the first case, we take a sparse interaction matrix in Eq.(\ref{eq:glv}), where $\bar{c}(N)$ is the probability of a non-zero entry. We then study the case when species interact on average with a core number of species which is independent of the size of the community, thereby $N\bar{c}(N)$ does not depend on $N$, at least for large $N$. The coefficients $\alpha_{ij}$ of pairs of interacting species are drawn from a distribution $\mathcal{Q}(\alpha_{ij})$, which is a function that does not depend on $N$ nor on $c$ (and $\alpha_{ii}=-1$). This assumption is not essential for making analytical progress, yet it introduces some elements of biological realism into the framework. In large communities, for instance, species that are distributed throughout space engage in local interactions with other species, irrespective of the system's spatial extent. Therefore this diluted version of the GLV system can integrate some spatial effects into the mean field model. Thus, we assume that $P_N(\alpha_{ij})=(1-\frac{c}{N})\delta(\alpha_{ij})+\frac{c}{N}\mathcal{Q}(\alpha_{ij})$, where $c $ is a positive constant ($\lim_N N\bar{c}(N)=c$) and $\delta(x)$ is a Dirac delta. If we substitute $P_N(\alpha_{ij})$ into Eq.(\ref{eq:distr eta}) and assume that $c\ll1$, we obtain at order $\mathcal{O}(c)$ (See SI for the exact equation satisfied by $P_\eta$)

\begin{equation}
    P_{\eta}(\eta)= (1-c)\delta(\eta)+c\mathcal{Q}(\eta)\quad,
    \label{rsa:dilute_first order}
\end{equation}
therefore the distribution of the population densities (SAD or RSA) at order $\mathcal{O}(c)$ reads (see also Eq.(\ref{eq:x eta}))

\begin{equation}
    Q_{x}(x)= (1-c)\delta(x-1)+c\delta(x)\int_{-\infty}^{-1}\de{\eta}\mathcal{Q}(\eta) + 
    c\mathcal{Q}(x-1)\Theta(x)\quad,
    \label{rsa:dilute}
\end{equation}
where $\Theta(x)$ is a Heaviside function. In the SI we have included the expression of $Q_{x}(x)$ at order $\mathcal{O}(c^2)$. Remarkably, we note that for $x>1$ Eq.(\ref{rsa:dilute}) provides a direct connection between the SAD -- a macroscopic pattern -- and the distribution of species interactions, which in principle can be empirically tested.

\begin{figure}[h!]
    \centering
    \includegraphics[width=8.cm]{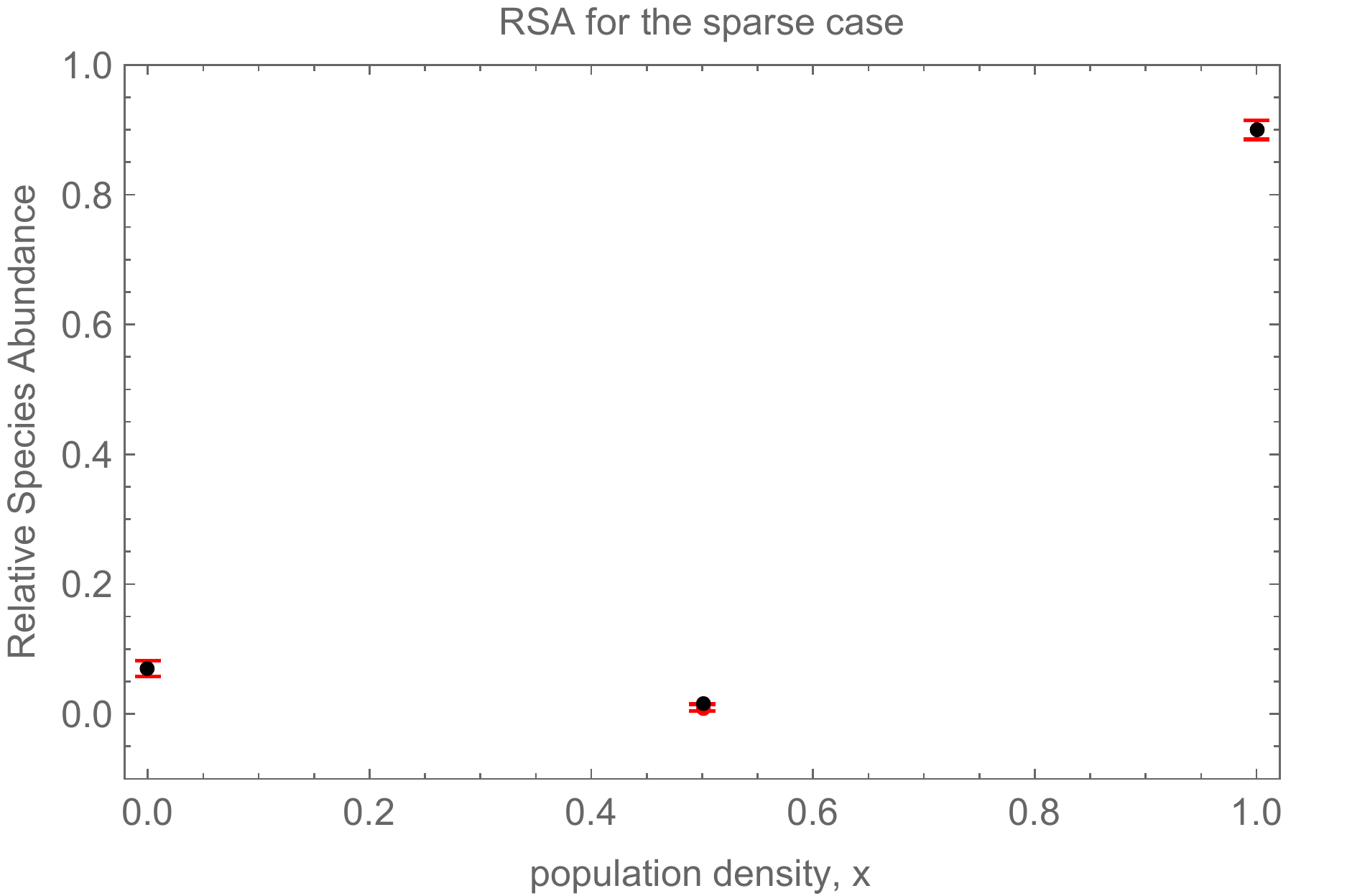}
    \includegraphics[width=8.cm]{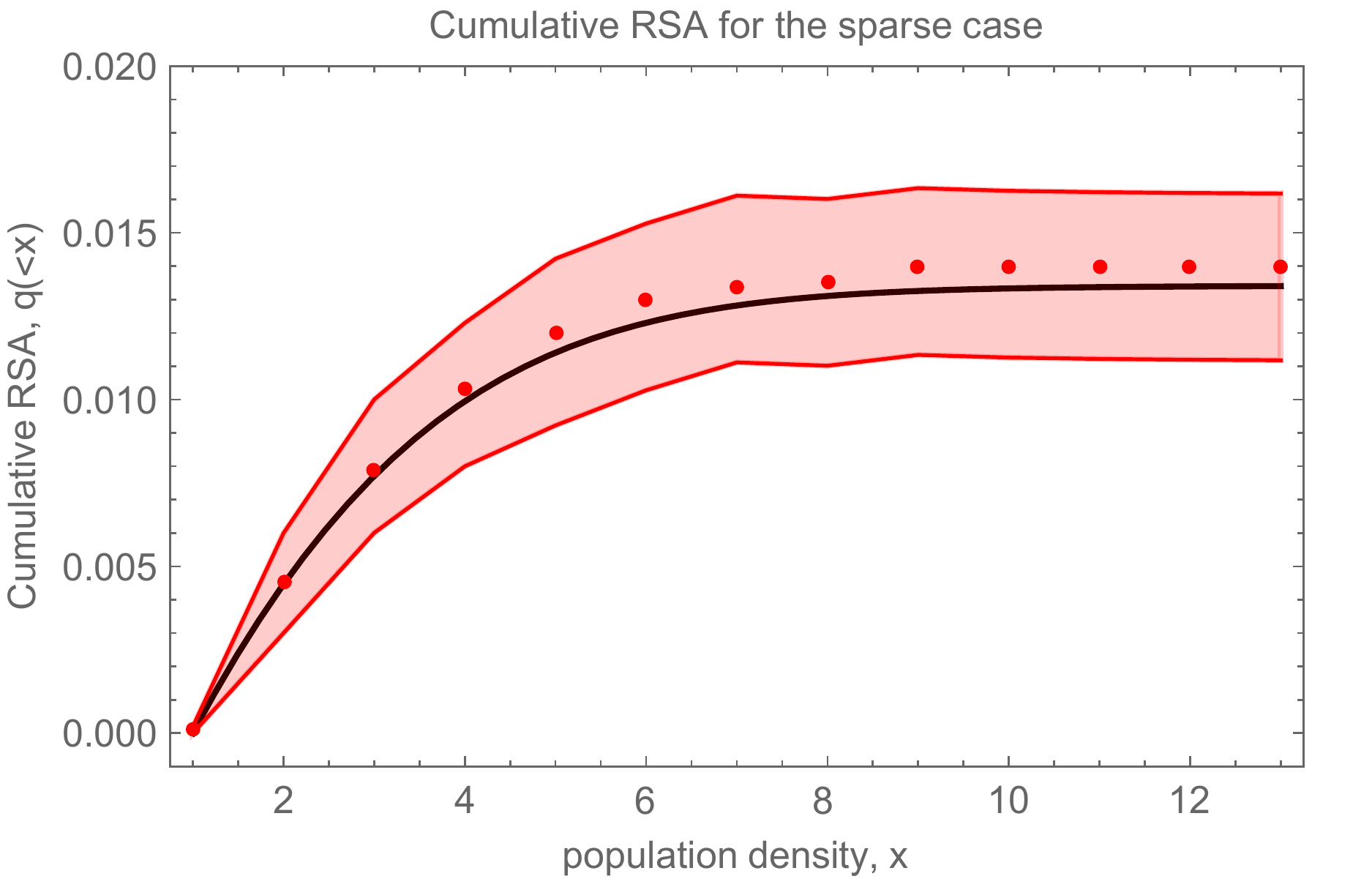}
    \caption{\textbf{The sparse case (small $c$)}. The \textbf{left panel} shows the Relative Species Abundance for species with population densities between 0 and 1. We run 200 realizations of a GLV system with 1,000 species until stationarity, in which the average number of interacting species was $\lim_N N\bar{c}(N)=c=0.1$. 
    The non-zero interaction coefficients in the GLV were drawn from a Gaussian distribution with mean $\mu=-4$ and standard deviation $\sigma=4$ (with no rescaling of the parameters). The black dot at $x=0$ represents the fraction of extinct species ($x<10^{-5}$) and analytically is given by $\frac{c}{2}(1-Erf[\frac{1+\mu}{\sqrt{2}\sigma}])$ (at order $\mathcal{O}(c)$); the black dot at $x=1/2$ represents the fraction of surviving species with $0<x<1$ and is given by $\frac{c}{2}(Erf[\frac{-\mu}{\sqrt{2}\sigma}]-Erf[\frac{-\mu-1}{\sqrt{2}\sigma}])$ (at order $\mathcal{O}(c)$); the black dot at $x=1$ represents the fraction of species with densities $1\leq x \leq 1.01$ and is given by $1-c$ (at order $\mathcal{O}(c)$). The red dots (not visible) represent the average fractions and the error bars give the first and third quartiles from the simulations, respectively. \\ 
    The \textbf{right panel} shows the fraction of species with a population density smaller than $x$ for species with densities larger than 1 for the same GLV system. The solid black line represents the analytical prediction calculated at order $\mathcal{O}(c^2)$ (see S.I.). The red dots represent the corresponding fraction from the simulations, the red strip gives the first and third quartiles from the simulations.}
    \label{fig:rsa_dilute}
\end{figure}

\begin{figure}[h!]
    \centering
    \includegraphics[width=10cm]{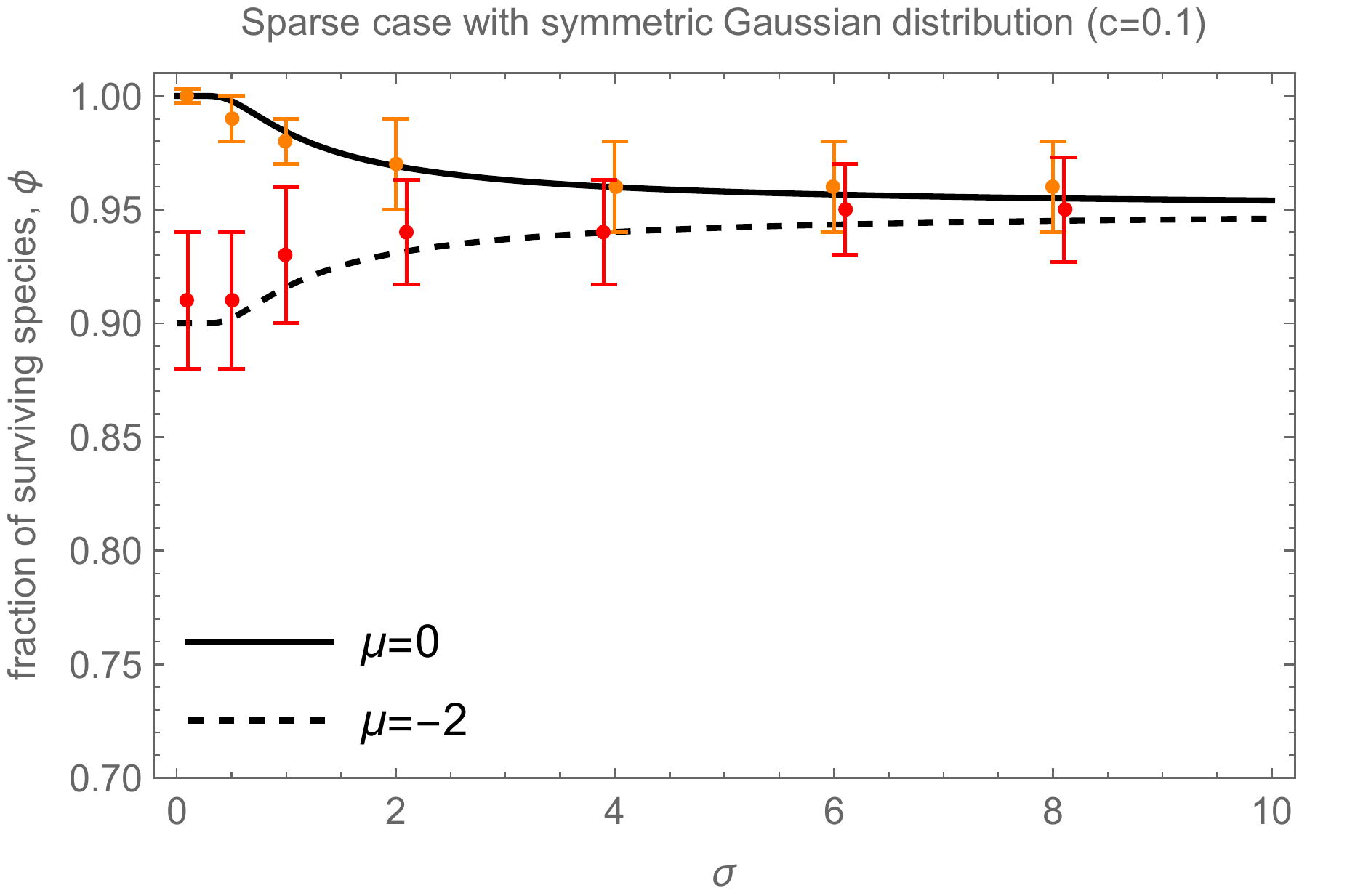}
    \caption{\textbf{The sparse case (small $c$).} Fraction of surviving species (population density $x>10^{-5}$) as a function of the standard deviation (with no rescaling). The interaction coefficients were drawn from a Gaussian distribution with mean $\mu=0,-2$. Solid and dashed lines come from the analytical predictions $\phi(\sigma)=1-c+\frac{c}{2}(1+Erf(\frac{1+\mu}{\sqrt{2}\sigma}))$ (at order $\mathcal{O}(c)$) for $\lim_N N\bar{c}(N)=c=0.1$, which was obtained from the integration of Eq.(\ref{rsa:dilute}). Points represent averages and error bars a standard deviation from 50 realizations of a system with 100 interacting species run until stationarity.}
    \label{fig:fracsurv_sparse}
\end{figure}

We have compared the analytical predictions against the simulations of the sparse GLV. We have derived the predictions of the theory for the RSA of species with population densities $0\leq x\leq 1$ and those with $x\geq 1$. Fig.(\ref{fig:rsa_dilute}) shows a comparison between the simulations for relatively small $c$ and the fraction of species with a population density smaller than $x$, calculated at order $\mathcal{O}(c^2)$ (see SI). The agreement is very good and is also confirmed between the analytical prediction of the fraction of surviving species and the simulation of a GLV system with a connectivity $c$ far below the percolation threshold (see Fig.(\ref{fig:fracsurv_sparse})).

\begin{figure}[h!]
    \centering
    \includegraphics[width=8.cm]{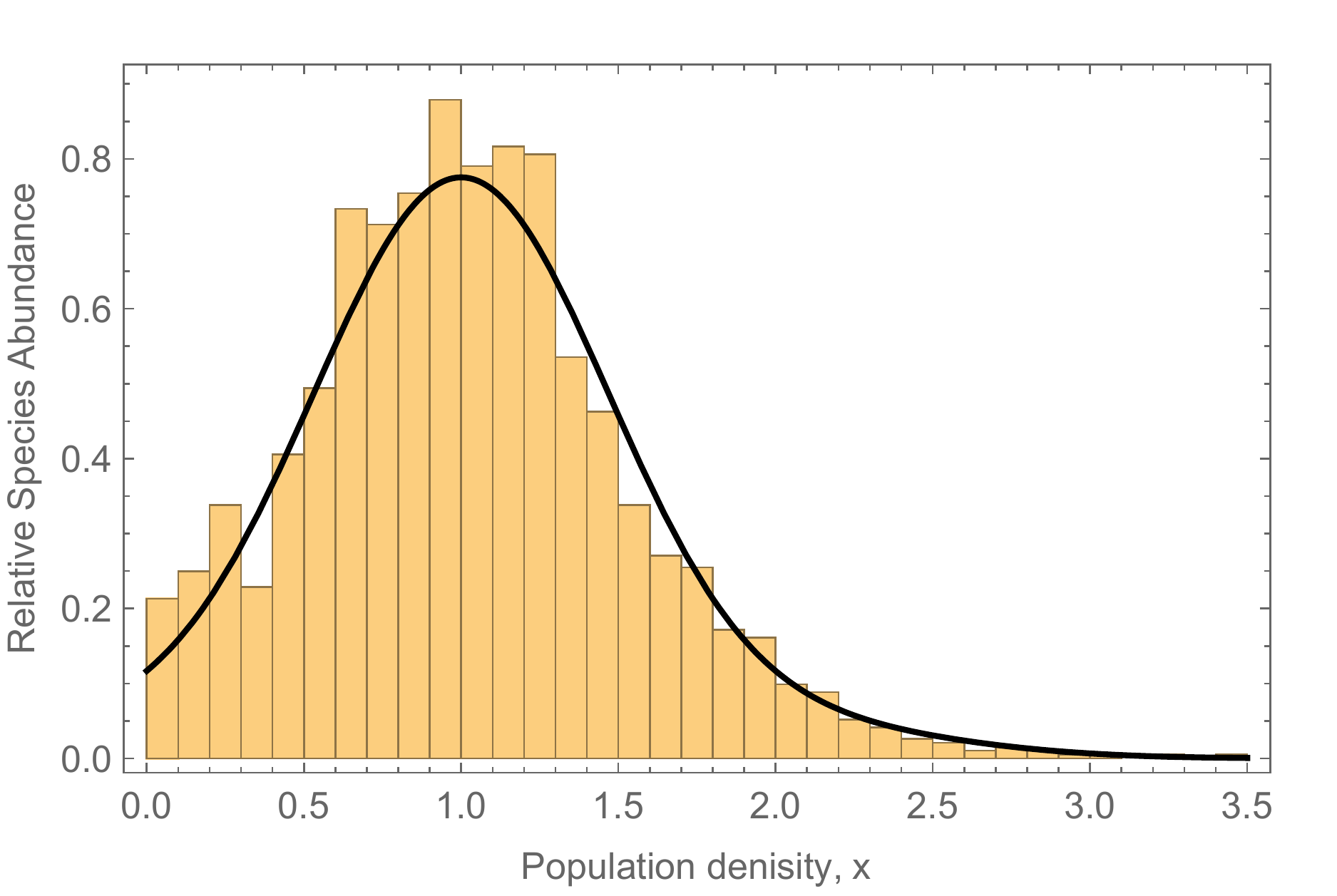}
    \includegraphics[width=8.cm]{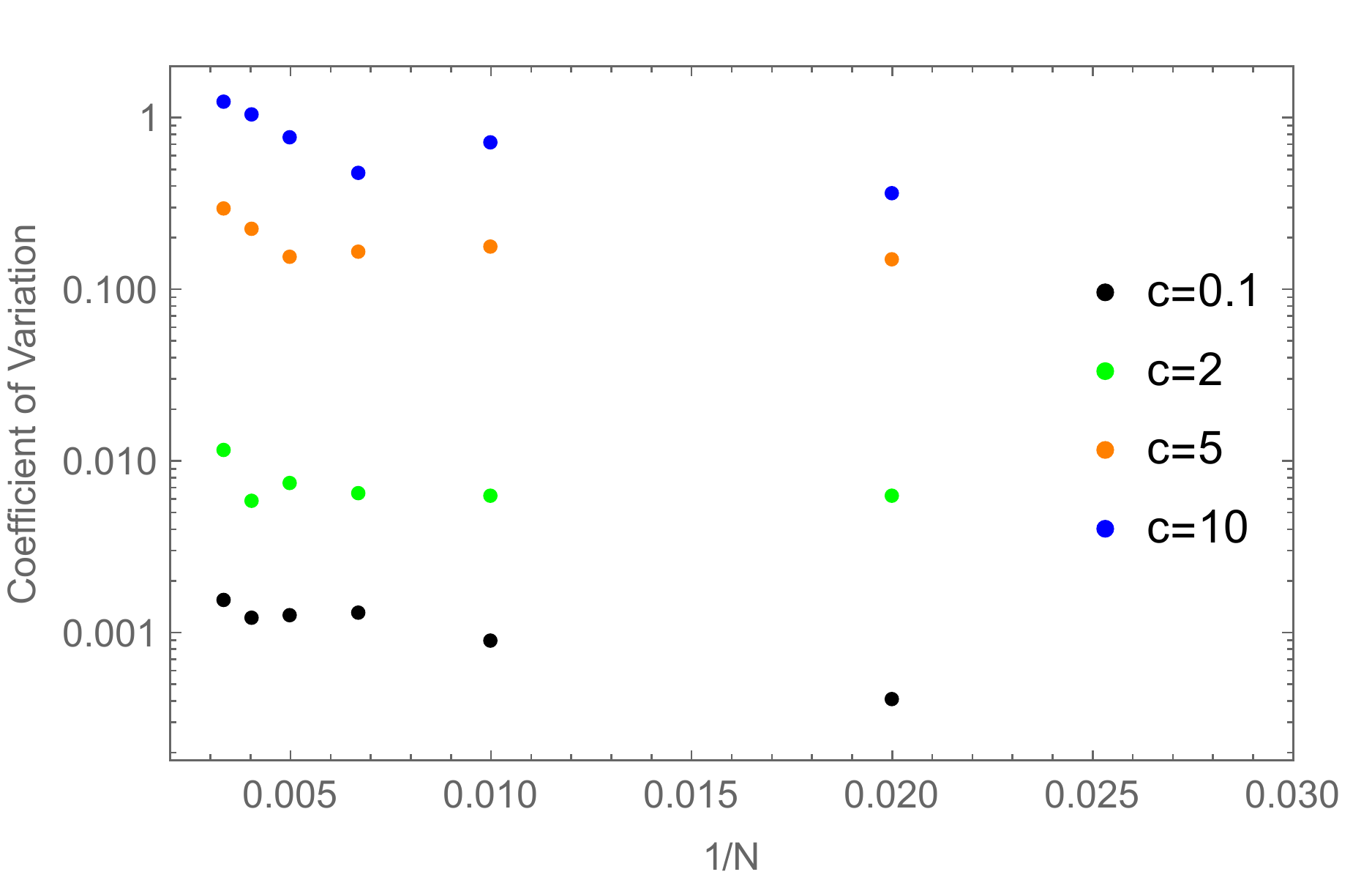}
    \caption{\textbf{The sparse case (large $c$)}. The \textbf{left panel} shows the analytic curve (solid black line) for the relative species abundance obtained from Eq.(\ref{eq:x eta}), expanding Eq.(\ref{eq:distr eta}) at order $\mathcal{O}(c^{-1})$ for large $c$ and assuming that $F_c(z)=c(\exp[\phi(z)/c]-1)$. The figure refers to the Gaussian case, namely, $\phi(z)=-\sigma^2 z^2/2$, where  $\sigma=0.5$ and $\lim_N N\bar{c}(N)=c=6$. The histogram represents the corresponding RSA for the non-extinct species obtained from the numerical simulation of the GLV equations, namely Eq.(\ref{eq:glv}), with the same parameters and $N=2,000$.\\ 
    The \textbf{right panel} shows the Coefficient of Variation (standard deviation to mean ratio) as a function of $N^{-1}$, being $N$ the number of interacting species ($N=50,100,150,200,250,300$); as before $c$ is the average number of interacting species. The interaction coefficients in the GLV were drawn from a Gaussian distribution with mean $\mu=-1$ and standard deviation $\sigma=0.5$ (no rescaling of the parameters). Notice the emergence of large fluctuations of the population densities for large $c$ at stationarity, although the parameters $\mu$ and $\sigma$ lie in the region of single fixed points of the standard DMFT \cite{bunin2017ecological,galla2018dynamically}.}
    \label{fig:largeC}
\end{figure}

In the second scenario of dilution, every species interacts with an average number of species $c$ which is much larger than one, although they are not connected with all the species in the community as in the mean-field case. We still assume that $P_N(\alpha_{ij})=(1-\frac{c}{N})\delta(\alpha_{ij})+\frac{c}{N}\mathcal{Q}_c(\alpha_{ij})$ (and $\alpha_{ii}=-1$), but since
in the large $c$ limit we expect to recover the mean-field results of the previous section, we stipulate that $\mathcal{Q}_c(\alpha_{ij})$ is now a function that does depend on $c$. This assumption translates into a function $F_c(z)$ (defined in Eq.(\ref{eq:scaling inter})) that depends on $c$ and $\lim_{c}F_c(z)=\phi(z)$, where $\phi(z)$ is the function describing the mean-field case. A simple form that leads to the appropriate limits is $F_c(z)=c(\exp[\phi(z)/c]-1)$ (See SI). This scaling of $F_c(z)$ allows to expand Eq.(\ref{eq:distr eta}) in powers of $c^{-1}$ for large $c$, thus finding an approximate solution. In turn, this solution provides a way to obtain the relative species abundance as given from Eq.(\ref{eq:x eta}). In the SI we have included the expression of $Q_{x}(x)$ at order $\mathcal{O}(c^{-1})$, which we show in the left panel of Fig.\ref{fig:largeC}. The comparison with the numerical simulations of the GLV equations is very good, even for relatively small $c$. The right panel of Fig.\ref{fig:largeC} shows, instead, that the sparse GLV equations without an appropriate scaling of the interactions may reach configurations that are not described by the standard phase diagram of the fully connected case even for large $c$.

\section{Discussion and Conclusions}
The dynamical mean-field theory has been recently applied to study large ecosystems driven by generalized Lotka-Volterra equations, in which the interactions between species are modeled using random couplings. It represents a natural generalization of the conceptual framework pioneered by May in his seminal 1972 paper and as such it has allowed the investigation of several features of disordered ecosystems. As the complexity (i.e, the total number of species) is large, the population densities of species may reach stable fixed points, they may fluctuate wildly yet within bounded domains, or go unbounded when the interaction coefficients vary too much among pairs of species. The stationary distribution of population densities is rarely different from a truncated Gaussian, which is also far from the empirical distributions found in real ecosystems, either in microbial communities or in large scale ecosystems. This lack of flexibility is closely related to the \textit{universality} of the SAD obtained in the standard DMFT, where the ditribution of species' interactions depends only on the first two cumulants. Within this framework the emergent SADs are very similar in shape, because what ultimately matters is the mean and variance of those couplings. In this regard, the SAD does not depend on the fine details of how species interact and therefore it would be impossible to reconstruct the structure of interactions.

On the contrary, our approach generalizes the standard DMFT and by doing that, it becomes more sensitive to the heterogeneity of the couplings. This breakdown of universality is informative, as it gives us an opportunity to develop a deeper understanding of the underlying species' interactions. The SAD does no longer depend on the first two cumulants only, but on all cumulants of the distribution of species' couplings. This increased level of information allows us to build up a link between the SAD and species interactions, which can now be probed by a macroscopic pattern, a much more accessible tool. Furthermore, this framework suggests that the mean and variance of interactions may be no longer sufficient to describe the stability of a large interacting system.

These consequences and comments are not limited to ecological communities as our version of the DMFT can be applied to a variety of other situations, including glassy systems and neural networks. It will be of the utmost interest to understand how much of this framework will be confirmed empirically.

\begin{acknowledgments}
S.A. and A.M. very much thank Giacomo Gradenigo, Gabriele Manganelli, Giorgio Nicoletti and Samir Suweis for insightful discussions. S.A. and A.M. also acknowledge the support of the NBFC to the University of Padova, funded by the Italian Ministry of University and Research, PNRR, Missione 4, Componente 2, ``Dalla ricerca all’impresa'', Investimento 1.4, Project CN00000033. 
\end{acknowledgments}

\bibliography{sn-bibliography}


\end{document}